\documentclass[twocolumn]{article}

\usepackage{fullpage}
\usepackage{graphicx}
\usepackage{epstopdf}
\usepackage{amsmath}
\usepackage[pagewise, switch]{lineno}
\usepackage{authblk}
\usepackage{xcolor}
\usepackage{array}

\begin{document}

	\title{X-ray photodesorption from water ice in protoplanetary disks and X-ray dominated regions} 

	 \author[1]{R. Dupuy\footnote{E-mail: remi.dupuy@obspm.fr}}
	 \author[1]{M. Bertin}
	 \author[1]{G. F\'{e}raud}
	 \author[1]{M. Hassenfratz}
	 \author[1]{X. Michaut}
	 \author[1]{T. Putaud}
	 \author[1]{L. Philippe}
	 \author[1]{P. Jeseck}
	 \author[2]{M. Angelucci}
	 \author[2]{R. Cimino}
	 \author[3]{V. Baglin}
	 \author[4]{C. Romanzin}
	 \author[1]{J.-H. Fillion} 
	 
	 \affil[1]{Sorbonne Universit\'e, Observatoire de Paris, Universit\'e PSL, CNRS, LERMA, F-75005, Paris, France}
	 \affil[2]{Laboratori Nazionali di Frascati (LNF)-INFN I-00044 Frascati}
	 \affil[3]{CERN, CH-1211 Geneva 23, Switzerland}
	 \affil[4]{Laboratoire de Chimie Physique, CNRS, Univ. Paris-Sud, Universit\'e Paris-Saclay, 91405, Orsay, France}

	\date{}
	
	\maketitle

\textbf{Water is the main constituent of interstellar ices, and it plays a key role in the evolution of many regions of the interstellar medium, from molecular clouds to planet-forming disks \cite{vandishoeck2014a}. In cold regions of the ISM, water is expected to be completely frozen out onto the dust grains. Nonetheless, observations indicate the presence of cold water vapor, implying that non-thermal desorption mechanisms are at play. Photodesorption by UV photons has been proposed to explain these observations \cite{hogerheijde2011, caselli2012}, with the support of extensive experimental and theoretical work on ice analogues \cite{arasa2015, munozcaro2016, bertin2016}. In contrast, photodesorption by X-rays, another viable mechanism, has been little studied. The potential of this process to desorb key molecules, such as water, intact rather than fragmented or ionised, remains unexplored. We experimentally investigated X-ray photodesorption from water ice, monitoring all desorbing species. We find that desorption of neutral water is efficient, while ion desorption is minor. We derive for the first time yields that can be implemented in astrochemical models. These results open up the possibility of taking into account the X-ray photodesorption process in the modelling of protoplanetary disks or X-ray dominated regions.}

\begin{figure}
    \includegraphics[trim={3.5cm 5cm 7.5cm 3.5cm},clip,width=\linewidth]{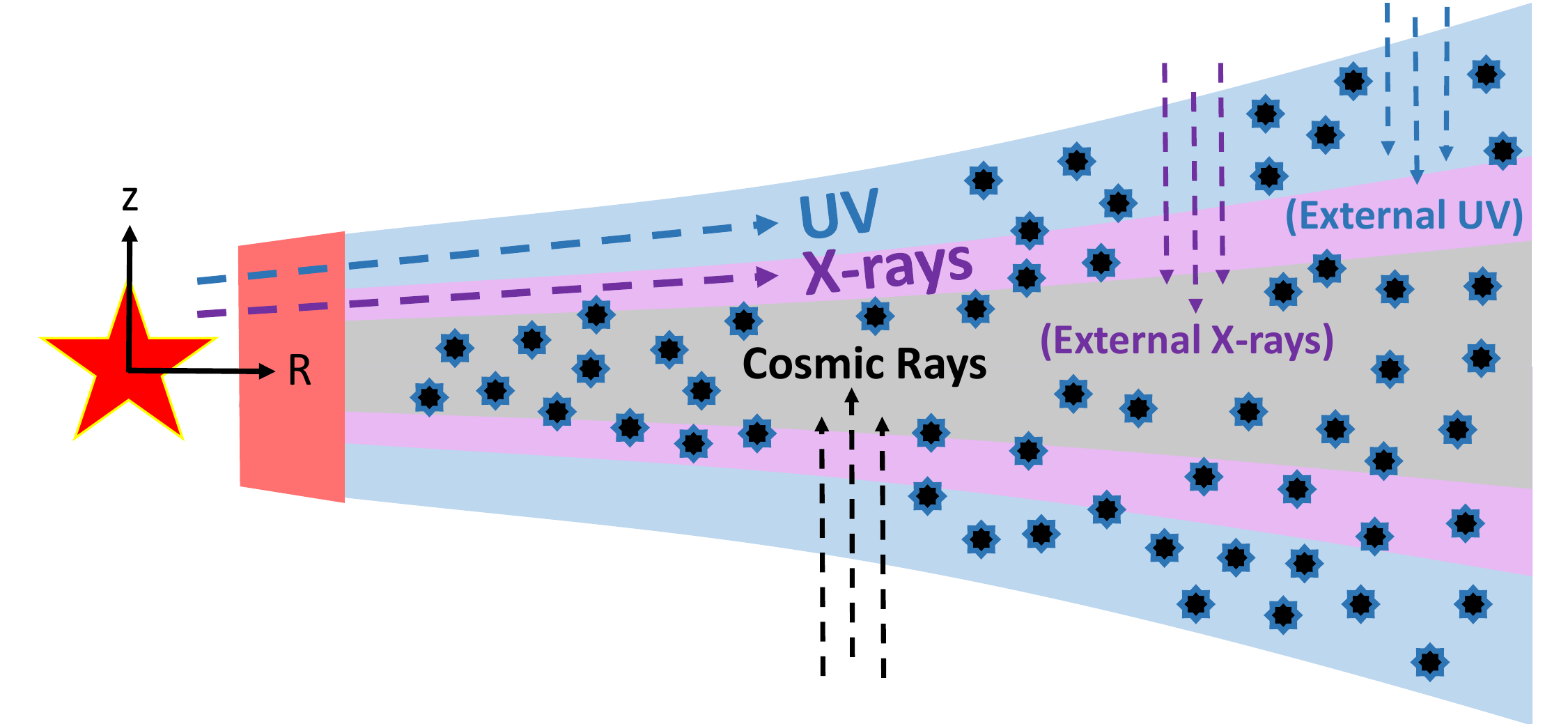}
    \caption{Schematic representation of a vertical slice of a protoplanetary disk showing the various sources of irradiation in the ice-containing regions. To the left is the central star, with the red zone being a hot region of dust sublimation. Icy grains are symbolized by black and blue dots. The limits between the different regions (UV-dominated, X-ray dominated or CR-dominated, as well as the "snow surface" locating the onset of ices) are all disk-dependent, see the text for more details. The outer part of the disk can be illuminated by external UV photons from the ISRF, but also by external X-rays \cite{rab2018}.}
    \label{sketch}
\end{figure}

Numerous studies have explored the effects of X-rays on the chemistry of various regions of the ISM \cite{aikawa1999, walsh2012, maloney1996, meijerink2007}, with a few taking into account solid phase processes \cite{walsh2010, stauber2006}. In the example of protoplanetary disks, as sketched in figure \ref{sketch}, the X-ray dominated region corresponds to a thin layer. The extent of the X-ray layer will depend on the type of star, the degree of evolution of the disk (dust settling, etc) and the physical model of disk considered. In the T Tauri model of ref \cite{walsh2010}, the limit between the UV and X-ray dominated layers occurs around z/R $\sim$ 0.2. The location of ices is disk-dependent as well, with the horizontal onset varying between $<$1 and 10 AU. Beyond a few tens of AU dust temperatures are cold enough for water to freeze out throughout the whole disk, and the X-ray layer overlaps with the icy region (cf fig.\ref{sketch}). The region where X-rays are dominant can extend to the whole midplane if the disk is shielded from cosmic rays, as suggested by \cite{cleeves2015a}.

The goal of this study is to provide experimental constraints on X-ray photodesorption, which allows modellers to assess the relevance of this process to the physics and chemistry of disks and other regions. Previous experimental studies have mainly focused on ion desorption \cite{rosenberg1983, coulman1990, mase2003}, and only a few have attempted to derive quantitative desorption yields \cite{pilling2012}. We used synchrotron radiation from the SEXTANTS beamline (SOLEIL facility) to irradiate amorphous solid water at either 15 K or 90 K in an ultra-high vacuum chamber (see Methods). Our set-up allows us to monitor directly in the gas phase the desorption of neutrals, cations and anions through mass spectrometry. Measuring desorption in the gas phase allows to unambiguously identify which species are desorbing, and prevents overestimation of the rates due to photochemistry into the ice \cite{bertin2016}. We are thus able to obtain a complete picture of the desorbing species, within our limits of sensitivity. The photodesorption yield (Y$^{inc}$; number of species desorbed per incident photon) at 550 eV of the species observed is given in table 1. One of our main findings is that the desorption of neutral species (H$_2$O, O$_2$ and H$_2$) is one to two orders of magnitude higher than the desorption of H$^+$, the most abundant ion. Therefore the investigation of neutral desorption is crucial for a quantitative understanding of X-ray photodesorption. This letter will focus on neutral desorption, while more information about ion desorption can be found in the Supplementary Information.

\begin{figure*}[ht]
    \includegraphics[trim={1.5cm 1cm 1.5cm 1.5cm},clip,width=\linewidth]{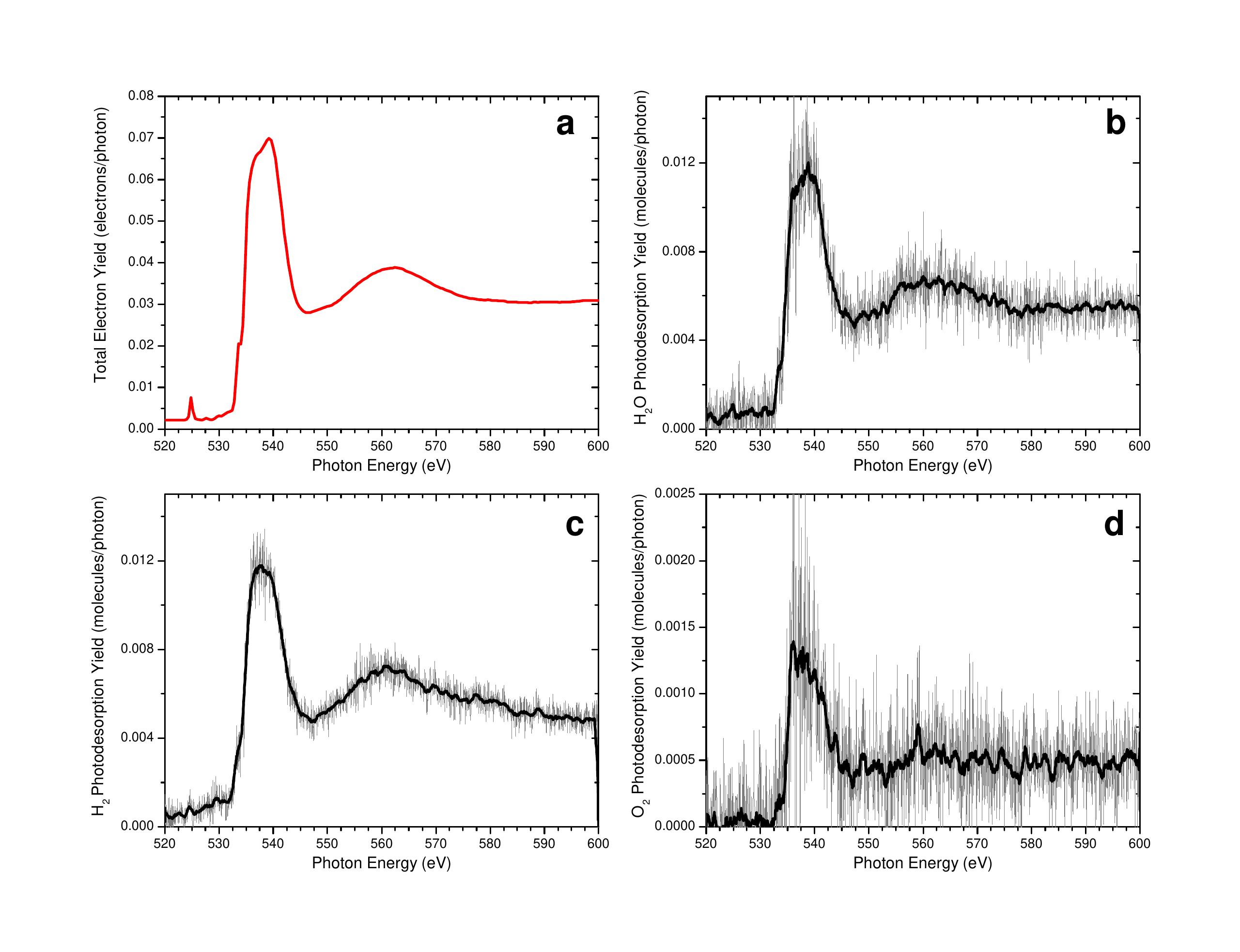}
    \caption{Photodesorption yields of the neutral species observed, as a function of photon energy. a. Total electron yield in electron per incident photon, and b. H$_2$O, c. H$_2$ and d. O$_2$ photodesorption yields in molecules per incident photon, as a function of photon energy, for a 100 ML compact amorphouse solid water ice at 15 K. The grey traces are the raw data while the black traces have been smoothed. The total electron yield is assumed to represent the absorption of the ice (see Methods).}
    \label{spec}
\end{figure*}

Figure \ref{spec}b shows the photodesorption yield of H$_2$O at 15 K and figure \ref{spec}a the total electron yield (which is assumed to reflect the absorption of the ice, see Methods) as a function of photon energy, between 520 and 600 eV. Detailed interpretations have been made of the X-ray absorption spectrum of ice between 530 and 550 eV \cite{nilsson2010}. The molecular orbitals of water are strongly affected by hydrogen bonding in the condensed phase, and thus the different features cannot simply be seen as inherited from the orbitals of the free molecules. However, a distinction can be made between subsets of molecules with different coordinations: the pre- and main-edge (533.6 and 535 eV) are mostly from molecules with a broken or distorted donor H-bond while the post edge corresponds more to fully coordinated molecules. The 561 eV feature is the first EXAFS oscillation \cite{parent2002}, and small features before 533 eV are attributed to photoproducts in the ice \cite{laffon2006}. The latter is discussed in more details later in the text. The photodesorption yield of H$_2$O follows the total electron yield, which is evidence that photodesorption is indeed caused by excitation or ionisation of the O 1s electron of the ice bulk water molecules, and not by the substrate.

   \begin{table*}
      \caption[]{Photodesorption yields (Y$^{inc}$) at 550 eV for a compact amorphous solid water ice at either 15K or 90K (desorbed species per incident photon)}
         \label{table1}
         \def\arraystretch{1.2}
         \begin{tabular}{c c c c c c c c c}
          & \multicolumn{2}{c}{Neutrals} & & \multicolumn{2}{c}{Cations} & & \multicolumn{2}{c}{Anions} \\
                    & 90 K & 15 K & & 90 K & 15 K & & 90 K & 15 K \\
                    \hline
             H$_2$O & $3.8 \times 10^{-3}$ & $3.4 \times 10^{-3}$ & H$^+$             & 
$1 \times 10^{-4}$   & $4.3 \times 10^{-5}$ & H$^-$   & $1.3 \times 10^{-5}$ & $6.5 \times 10^{-6}$ \\
             H$_2$  & $8.9 \times 10^{-3}$ & $5.3 \times 10^{-3}$ & H$_2^+$           &
$5 \times 10^{-7}$   & $1.6 \times 10^{-7}$ & H$_2^-$ & $3 \times 10^{-10}$  & 						\\
             O$_2$  & $2.5 \times 10^{-3}$ & $4 \times 10^{-4}$   & H$_3^+$           &
$2.5 \times 10^{-9}$ & 						& O$^-$   & $1.3 \times 10^{-7}$ & $2.3 \times 10^{-8}$ \\
             OH     & $< 1 \times 10^{-3}$ & $< 1 \times 10^{-3}$ & O$^+$             &
$3.6 \times 10^{-8 }$ & $2.8 \times 10^{-8}$ & OH$^-$  & $2.2 \times 10^{-8}$ & $6.4 \times 10^{-9}$ \\ 
             		& 					   & 					  & OH$^+$            &
$6 \times 10^{-9}$   & $4.6 \times 10^{-9}$ & H$_2$O$^-$ & $3 \times 10^{-10}$ & 					\\
             		& 					   & 					  & H$_2$O$^+$        &
$3 \times 10^{-9}$   & $3.7 \times 10^{-9}$ & O$_2^-$ & $8 \times 10^{-10}$  & 						\\ 
            		& 					   & 					  & H$_3$O$^+$        &
$1.2 \times 10^{-8}$ & $1.6 \times 10^{-8}$ & 		  & 		 			 & 						\\ 
             		& 					   & 					  & O$_2^+$           &
$2.9 \times 10^{-8}$ & 					    & 		  & 					 & 						\\ 
             		& 					   & 					  & (H$_2$O)$_2$H$^+$ &
$1.9 \times 10^{-8}$ & $2.4 \times 10^{-8}$ &		  & 					 & 						\\    
         \end{tabular}
   \end{table*}

We now consider the number of molecules desorbed per absorbed photon (Y$^{abs}$), which is a quantity more intrinsic to the efficiency of the X-ray photodesorption process. It can be derived knowing the absorption cross-section of water and the number of monolayers below the surface involved in the photodesorption process. The former requires approximating the unknown solid-phase water absorption to the gas-phase one. The latter requires consideration of the length scale of the various processes involved. A rough estimate is Y$^{abs}$ $\sim$ 0.2 H$_2$O molecule desorbed per absorbed photon for a thickness of 50 ML (see Methods).

Following a core electronic transition, energy is released through Auger decay, where most of the energy is carried away by an electron. In water this electron has 500 eV of kinetic energy and will create multiple secondary valence ionisation and excitation events on its track, leading to desorption of molecules, which is usually termed X-ray induced electron-stimulated desorption (XESD). As most of the energy goes into the Auger electron, XESD will usually dominate the desorption yield. The fact that the desorption yield of neutrals is much higher than that of ions is probably due in part to the inability of secondary electrons to produce ions with sufficient kinetic energy to desorb and overcome interactions specific to charged particles (stabilisation by polar molecules, reneutralisation...). The value we derived earlier for Y$^{abs}$ is close to the yield derived for 87 eV electron-stimulated desorption, 0.5 molecule per electron \cite{petrik2005}, which is another indication that XESD dominates. It is much higher than the one derived for UV photodesorption, $\sim 10^{-2}$ molecules per absorbed photon \cite{arasa2015}. However, it appears that this quantity scales roughly with the deposited energy, as X-ray photons carry about two orders of magnitude more energy than UV photons. 

\begin{figure}
    \includegraphics[trim={1.5cm 1cm 1.5cm 1.5cm},clip,width=\linewidth]{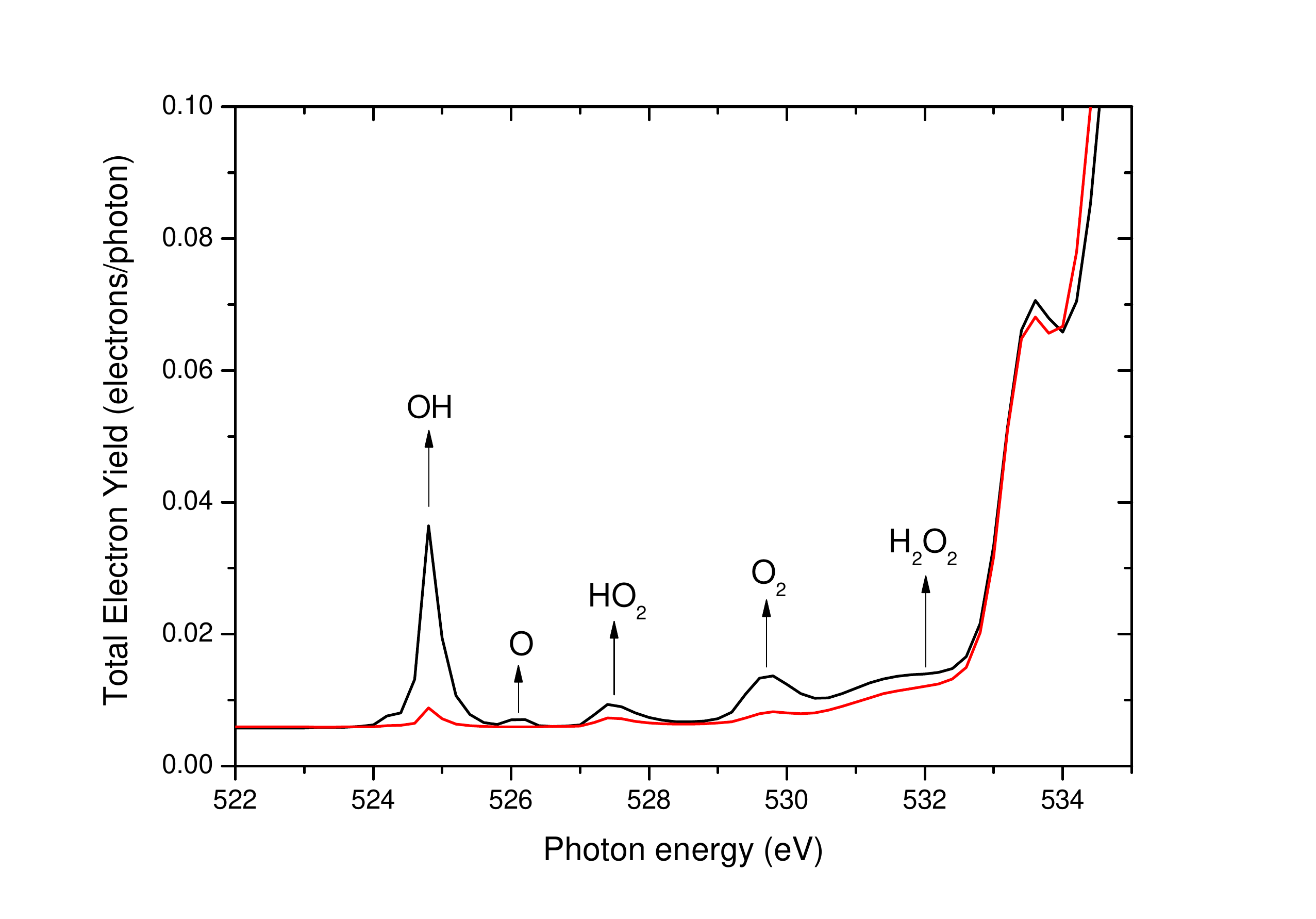}
    \caption{Total electron yield in electron per incident photon, as a function of photon energy. The data is for a 100 ML c-ASW at 15 K (black trace) and 90 K (red trace). The attribution of the peaks follows ref. \cite{laffon2006}.}
    \label{preedge}
\end{figure}

We also observe photodesorption of other neutral species, O$_2$ and H$_2$ (see fig \ref{spec}c, \ref{spec} and table 1) with a spectral behaviour similar to H$_2$O. The desorption of these species as well as others such as OH fragments has been studied in detail in UV  irradiation experiments \cite{yabushita2013}. We do not observe the desorption of OH radicals, and the estimated sensitivity limit, considering the experimental noise on this channel, is 10$^{-3}$ molecules per incident photon. We do not observe desorption of HO$_2$ or H$_2$O$_2$ either. The desorption of H$_2$ and O$_2$ is an indicator of the chemistry occurring in the ice. This chemistry can also be probed in situ through the aforementioned small features before 533 eV in the absorption spectrum. Fig. \ref{preedge} is a zoom into the total electron yield between 522 and 535 eV. Several peaks are seen that can be attributed to species other than H$_2$O, produced by the irradiation. Following the attributions of ref \cite{laffon2006}, we can identify OH, O, HO$_2$, O$_2$ and H$_2$O$_2$. There is a noticeable difference between the spectra at 15K and 90K: at 90K radical species such as OH and O diffuse and thus are consumed, which means that they are much less abundant. O$_2$ diffuses as well and can thermally desorb, so that its abundance in the ice is lower. The abundance of H$_2$O$_2$, on the other hand, while difficult to estimate, stays approximately constant. There is no evolution of the observed photoproducts with photon dose. All the data shown here thus correspond to a steady state, where formation and destruction of the species has reached an equilibrium. The sum of the photoproducts does not exceed a few percent of the total content of the ice \cite{laffon2006}, showing the resilience of pure water ice to X-ray irradiation. 

The photodesorption yield of ices that have been exposed to much lower photon doses and where this steady state has not been reached, as may be the case in the ISM, could be different for O$_2$ and H$_2$, as chemistry plays an important role. It should not differ much for water, as its desorption is presumably dominated by single-excitation processes (see also Methods).

The observed temperature effects for desorbing O$_2$ and H$_2$ (Table 1) are similar to those observed in the UV range \cite{cruz-diaz2018}, and are well explained by diffusion and thermal desorption processes. At 15 K O$_2$ is not mobile and thus its desorption yield is very low and confined to the molecules produced near the surface. At 90 K, O$_2$ can diffuse and thermally desorb once it is formed, and thus its desorption is increased by an order of magnitude. H$_2$ can diffuse and thermally desorb at both temperatures, but its diffusion and desorption are higher at 90 K. The enhancement of the thermal diffusion of radicals at 90 K, as seen in the absorption spectrum (fig. \ref{preedge}), can also participate to the observed effects. A small temperature effect is also observed for H$_2$O desorption, as was found in the UV range \cite{arasa2010}. This behaviour suggests that H$_2$O desorption is not linked, or only in a minor part, to diffusion processes and thermal desorption.

 	\begin{table*}
	\begin{minipage}{17cm}
	\centering
	\caption[]{Average photodesorption yields of intact water (H$_2$O molecules per incident photon) for the X-ray spectra of different regions at different attenuations\footnotemark[1]. The cumulative uncertainties of the various sources of errors are estimated to be $\pm$ 50\%, as detailed in Methods.}
	\def\arraystretch{1.2}
		\begin{tabular}{c c c}
				   				  & TW Hya (young star) & MKN231 (ULIRG) \\
		\hline
		Source spectrum           & $2.3 \pm 1.2 \times 10^{-3}$ & $1 \pm 0.5 \times 10^{-3}$	\\
	    n$_H$ = 10$^{21}$ cm$^2$  & $2.5 \pm 1.3 \times 10^{-3}$ & $8 \pm 4 \times 10^{-4}$		\\
	    n$_H$ = 10$^{22}$ cm$^2$  & $1.2 \pm 0.6 \times 10^{-3}$ & $1.8 \pm 0.9 \times 10^{-4}$	\\
	    n$_H$ = 10$^{23}$ cm$^2$  & $1.2 \pm 0.6 \times 10^{-4}$ & $2 \pm 1 \times 10^{-5}$	    \\
	    n$_H$ = 10$^{24}$ cm$^2$  & $1.2 \pm 0.6 \times 10^{-5}$ & $6.6 \pm 3.3 \times 10^{-6}$	\\
		\end{tabular}	
	\footnotetext[1]{See text and Methods. The yields used are those for a c-ASW ice at 15K.}
	\end{minipage}
	\end{table*} 

Our experimental data is limited to the 520 - 600 eV range, while X-ray spectra in the ISM typically span from 0.1 to at least 10 keV. Therefore in order to be included in astrochemical models, the photodesorption yields need to be extrapolated to higher photon energies. Photodesorption of neutrals is proportional to absorption, and absorption is still dominated by O 1s ionisation from 600 eV up until 10 keV. The extrapolation can therefore be done using the absorption cross-section of water, once again using the gas-phase value (see Methods). The extrapolated yields from 150 eV to 10 keV are available in the Supplementary Information. The local average photodesorption yield can then be calculated by multiplying the energy-dependent photodesorption yields by the local X-ray spectrum, which gives the photodesorption yield per average incident photon (see Methods). The X-ray spectrum at a given location can vary a lot depending on the column density of gas traversed, as the softer X-rays are more attenuated. In order to illustrate this dependence, we took observation spectra from two different sources, TW Hya \cite{nomura2007} as an example of young star that may illuminate a protoplanetary disk or a protostellar envelope, and MKN231 \cite{braito2004} as an example of an ultra-luminous infrared galaxy (ULIRG) with AGN and starbust components, and we calculated the spectrum for different column densities traversed. Then we calculated the photodesorption yield per average incident photon. The results are given in table 2. Numbers vary from $2.5 \times 10^{-3}$ to $6.6 \times 10^{-6}$, illustrating both the strong source and column density dependence of the efficiency of X-ray photodesorption. These numbers can be used as a rough approximation in models, or the photodesorption yield adequate to the modelled environment can be calculated using our extrapolated energy-resolved photodesorption yields (Supplementary Information). 

UV photodesorption is important in the current understanding of gas-phase water in cold regions of the ISM \cite{hogerheijde2011,caselli2012}, and the existing experimental constraints allow it to be taken into account in models. Our experimental results now open up the possibility of taking into account the X-ray photodesorption process in regions of the ISM where ices and X-rays co-exist. As previously mentioned, one such region is the layer of protoplanetary disks where X-rays can dominate over UV photons and cosmic rays (fig.\ref{sketch}). One key element of the understanding of planet formation is the determination of the water snowline, which requires accurate modelling of all processes affecting gas phase and solid phase water in disks. Taking one specific example with the model of a T Tauri star presented in ref \cite{walsh2010}, the UV field is completely attenuated below z/R $\sim$ 0.2. The X-ray flux in this region varies between 10$^{-6}$-10$^{-4}$ erg/s/cm$^2$, i.e. 10$^3$-10$^5$ photons/s/cm$^2$. On the other hand, secondary UV produced by cosmic rays amount to $\sim$ 10$^3$ photons/s/cm$^2$ \cite{cecchi-pestellini1992} while direct cosmic ray sputtering has about the same efficiency as photodesorption from these secondary UV \cite{dartois2015a}. Considering the efficiency of X-ray photodesorption is comparable to UV photodesorption (10$^{-4}$-10$^{-3}$ H$_2$O molecules/incident photon), X-ray photodesorption is expected to play a role in this case. If we consider instead that cosmic rays are excluded from the midplane of the disk \cite{cleeves2015a}, then X-rays largely dominate. The relevance of X-ray photodesorption for a given disk, however, depends on many parameters as discussed before: the type of star, the X-ray spectrum, the physical model considered, etc. The case of M dwarf stars could be particularly interesting as they feature weak UV and strong X-ray emission, as well as extended regions of cold dust temperatures \cite{walsh2015}. 

Molecular clouds exposed to intense X-ray sources and modeled by so-called X-ray Dominated Regions (XDR) codes are also of interest. They include the envelopes of young stellar objects (YSO) \cite{stauber2006} and molecular clouds close to active galactic nuclei (AGN) \cite{meijerink2007} or strong supernovae shocks \cite{maloney1996}. As X-ray photons can penetrate deep into clouds and do not heat grains very efficiently, these regions can host ices. XDR codes could also include the X-ray photodesorption process to assess its relevance to the physics and chemistry of these various environments.

\section*{Methods}

\subsection*{Experimental set-up}

Experiments were conducted in the SPICES 2 set-up, a recently upgraded version of the SPICES set-up. SPICES 2 is an ultra-high vacuum chamber (base pressure $\sim 2.10^{-10}$ mbar) equipped with a rotatable cryostat at the tip of which a sample holder is attached. The temperature of the samples can be controlled with a 0.1 K precision between 15 and 300 K. The experiments were carried on a copper substrate (polycrystalline Oxygen Free High Conductivity copper) which is isolated electrically, but not thermally, from the sample holder by a kapton foil. This allows to read the drain current of the substrate generated by the electrons escaping the surface following X-ray absorption, the so-called Total Electron Yield (TEY). The number of escaping electrons is proportional to the number of absorbed photons, thus the drain current is proportional to the absorption of the ice. The chamber is also equipped with a gas-dosing system described below and two quadrupole mass spectrometers (QMS), one (from Pfeiffer Vacuum) used for the detection of neutral species and the other (from Hiden Analytical) for the detection of positive and negative ions. 

\subsection*{The SEXTANTS beamline}

The SPICES 2 set-up was brought to the SOLEIL synchrotron facility and connected to the SEXTANTS beamline described in \cite{sacchi2013}. For these experiments we used photons in the 520 - 600 eV range. Settings were such that the resolution was $\Delta E = 150$ meV and the flux, as measured with a silicon photodiode, was approximately $1.4 \times 10^{13}$ photons.s$^{-1}$, with little variation except for a significant dip around 535 eV due to the O 1s absorption of oxygen pollution on the optics of the beamline.
The absorption spectra of ice that we obtain from our total electron yield measurements, once corrected from the photon flux, compare well with those found in the literature and obtained using many different techniques \cite{nilsson2010}, which makes us confident that our spectra are not distorted by the structure of the flux. 
The unfocused beam was sent at a 45$^{\circ}$ incidence on the surface, which resulted in a spot of approximately 0.1 cm$^2$.
All experiments were performed using the horizontal polarisation setting, which with a 45$^{\circ}$ incidence results in a 50 \% in plane and 50 \% out of plane polarisation at the surface.

\subsection*{Ice Growth}

Water (liquid chromatography standard from Fluka) was purified by a series of freeze-pump-thaw cycles and the vapor pressure was let into a gas injection line. The ice is grown on the substrate by approaching a dosing tube 1 mm close to it and by injecting water vapor into the tube using a micro-leak valve. This allows to grow a relatively thick ice ($\sim$100 equivalent bilayers, BLeq) without increasing the pressure in the chamber to more than a few 10$^{-9}$ mbar. The thickness of the ice can be calibrated by using the temperature-programmed desorption (TPD) technique as described in \cite{doronin2015}. All water ices were grown at 90 K, which results in a compact amorphous solid water (c-ASW) structure. The irradiation experiments were subsequently performed either at 90 K or after cooling the ice to 15 K. The thickness of the ice insures that the influence of the copper substrate on the desorption processes is negligible (see discussion of the length scales of the processes below), as confirmed by the fact that no important desorption signal is seen apart from the resonances of water (oxygen adsorbed on copper would have been observed around 530 eV.\footnote{NIST XPS database https://srdata.nist.gov/xps/}).
As mentioned in the text, all spectra shown correspond to a steady state, i.e. two consecutive scans give the same results. This steady state is reached immediately in our experiments. From the estimate of the photon dose required to reach a steady state of the ice bulk chemistry in ref \cite{laffon2006} (a few 10$^{15}$ eV/cm$^2$) it should indeed be reached in just a few seconds. The erosion of the ice due to photodesorption during irradiation is also small ($\sim$2 ML/hour at 600 eV). Consequently the ice sample was renewed every few hours, which did not cause reproducibility issues. Pure water ice is resilient to X-ray irradiation, as the modifications induced are small, a conclusion that was reached also for ion bombardment \cite{dartois2015a}.

In the conditions of the interstellar medium, the photon fluxes are much lower than the ones used in our experiments, which raises the question of whether our results can be extrapolated to these regions. Considering, for example, a flux of 100 photons/s/cm$^2$, the steady state dose would be reached in about 1200 years, which is small as compared to the lifetime of a protoplanetary disk. Whether the steady state dose will be reached will however still depend on the location of the ice, while the fact that a realistic ice contains not just water but also other species will change the nature of the steady state. The yield of water and electrons, which are presumably dominated by single-excitation processes, should not be affected much by the irradiation history of the ice. However it could be the case for the desorption of species like O$_2$ and H$_2$ which are linked with chemistry that requires multiple excitation events.

\subsection*{Calibration of the photodesorption yields}

The flux of molecules desorbed from the surface upon irradiation is proportional to the signal current measured on the QMS, once it has been substracted from the contribution of the background, i.e. the signal current when the irradiation is off. In order to obtain the proportionality factor between molecule flux and signal current we use the above-mentioned TPD technique. TPD allows to determine when one monolayer of a given molecule has been deposited on the surface \cite{doronin2015}. In the present case, this calibration has been done on CO, where determination of the monolayer is unambiguous since the monolayer and the multilayer contributions appear at clearly distinct temperatures. The TPD behaviour of H$_2$O, on the other hand, is particularly complex and thus not very suited for this method: there is no clearly distinguishable monolayer peak and the TPD follows a zero order kinetic even in the sub-monolayer regime. However, once the proportionality factor for a given molecule is known, it can be deduced for other molecules by correcting for the molecule-dependent factors, as explained in e.g. \cite{dupuy2017}. These factors are the differences in electron-impact ionisation cross-sections (taken from the literature \cite{orient1987, straub1996}), and the differences in transmission and detection efficiencies in the QMS, which depend on the mass of the molecule and have been calibrated for our QMS.

When a monolayer of CO is deposited, knowing the density of CO \cite{vegard1930} and the size of the surface (15 x 15 mm), we thus know the absolute number of molecules on the surface. During a TPD experiment, all of these molecules are released into the gas phase. Thus the integrated signal of the TPD experiment, once subtracted from the background contribution, is proportional to the integrated flux of molecules, i.e. the absolute number of molecules that were present on the surface. The ratio of these two known quantities gives us the proportionality factor for thermal desorption. We then assume that the proportionality for photodesorbed molecules is similar to that for thermally desorbed molecules. For example, this means that we must make the assumption that the angular and speed distribution of desorbed products are the same in both cases. This is reasonable in the case of neutral photodesorption from amorphous molecular solids, where we do not expect high-energy molecules or strong orientational effects. The validity of this method was tested against another calibration technique used in UV photodesorption experiments. In the UV range, the photodesorption of CO can be calibrated by monitoring CO loss in the ice using infrared spectroscopy, as detailed in \cite{fayolle2011} (such a technique cannot be used in irradiation experiments where chemistry occurs and molecule loss cannot be related simply to desorption). Both techniques give similar results within the 50\% estimated uncertainty for UV CO photodesorption. 

We can estimate the uncertainty of this method, taking into account the uncertainty on the monolayer calibration ($\pm 20\%$), the flux, the electron-impact ionisation cross-sections ($\pm 10\%$) and the apparatus function ($\pm 40\%$), which makes a total of $\pm 50\%$. The uncertainty of the apparatus function for H$_2$ is larger and not easy to estimate, therefore the H$_2$ absolute yields should be taken with caution. 

\subsection*{Estimation of photodesorption yields per absorbed photon (Y$^{abs}$)}

The photodesorption yield per absorbed photon Y$^{abs}$ is derived from the following formula \cite{cruz-diaz2014}:
\begin{linenomath} $$ Y^{abs} = \frac{Y^{inc}}{1-e^{-\sigma N}} $$ \end{linenomath}
Where $\sigma$ is the absorption cross-section and N is the column density of molecules which we consider to be involved in the photodesorption process. As far as we know, the absorption cross-sections of water ice in the soft X-ray range are not available in the literature. Only the gas-phase cross-sections are \cite{berkowitz2002}, and we thus assume that they can be used here. While the cross-sections have no reason to be similar between ice and gas in the edge region, off-resonance (at 600 eV) it is a reasonable assumption because the cross-section is then mostly the atomic O 1s ionisation cross-section and does not depend significantly on the molecule or its phase.
The column density of molecules involved in the photodesorption process, i.e. the number of monolayers below the surface that participate in desorption, has been well constrained in the case of UV photodesorption of CO \cite{bertin2012}. Desorbed CO molecules come from the surface, and the excitations that lead to desorption are localized in the top 3 monolayers.
The situation is rather more complex here because we need to consider primary X-ray excitations as well as secondary excitations that follow Auger decay processes, and the mechanisms through which these excitations lead to desorption. Let us review the various length scales involved. With an absorption cross-section of the order of $5 \times 10^{-19}$ cm$^2$, our 100 ML thick ice will only absorb $\sim$ 5\% of the incident photons. The primary excitations can thus take place over the whole ice, however only a fraction of them will lead to desorption, depending on their distance to the surface. The electron cloud created by an Auger electron in water has an approximate radius of 10 nm \cite{timneanu2004}, i.e. $\sim$ 30 ML, thus secondary excitations can be created near the surface from molecules 30 ML deep in the ice. Additionally, the desorption mechanisms that follow secondary excitations have their own length scales. The possible mechanisms of desorption of H$_2$O that have been discussed for UV excitation include kick-out by hot H fragments, chemical recombination \cite{andersson2008} and exciton-induced dipole reversal \cite{desimone2013}, to which we can add at least, due to the possibility of ionisation, recombination of H$_3$O$^+$ ions with electrons and subsequent desorption of the H$_2$O product \cite{kimmel1994}, and collision-induced desorption \cite{redlich2006}. These processes involve the scales of hot H atom mean free path, fragment diffusion, exciton diffusion, and proton diffusion along the H-bond network. These length scales are not well constrained. Molecular dynamics simulation yield a hot H atom mean free path of $\sim$ 8 \.{A} \cite{andersson2008}, which is small enough to be neglected. The experimental findings of ref \cite{petrik2005} suggest a characteristic migration scale of excitons of about 25 ML. Taking into account the radius of the electron cloud and the migration scale of excitons, we will thus consider that the first 50 ML are involved in X-ray photodesorption for water. Using this characteristic thickness we derive a yield of about 0.2 H$_2$O molecule desorbed per absorbed photon at 600 eV and at 15 K. This should be considered as an order of magnitude estimate, considering the uncertainties on the length scales and absorption cross-sections mentioned above.

\subsection*{Estimation of average photodesorption yields for astrophysical models (Y$^{avg}$)}

The experimental photodesorption spectrum in the 520-600 eV range follows the absorption spectrum of the ice, and above 600 eV the absorption is still dominated by O 1s ionisation, therefore the physical and chemical processes involved are the same. We can thus extrapolate the photodesorption yield above 600 eV if we know the absorption cross-section of water ice. As mentioned above, this is not the case but it can be safely extrapolated from the gas-phase value off-resonance. Before the edge, we assume that the photodesorption yield is constant from 150 to 520 eV and equal to the value measured at 520 eV. Such an assumption is certainly wrong, as the absorption cross-section of H$_2$O is not constant in this region. However in this region we cannot make an extrapolation similar to the one described previously, since the cross-section is dominated by the continuum of the (2a$_1$)$^{-1}$ ionisation and other higher energy states, and those states are very different from the O 1s core hole state: they do not lead to an Auger decay. Changing the value of the photodesorption yield below 520 eV only leads to negligible changes in the average value, therefore this approximation, although crude, is not critical. The yields extrapolated from 150 eV to 10 keV are available in the Supplementary Information, along with the values of the absorption cross-section used.

The extrapolated photodesorption yields are then multiplied by a normalized X-ray spectrum. In order to illustrate the average yields obtained for various relevant regions at different attenuations, we retrieved observed X-ray spectra of TW Hya \cite{nomura2007} and of the ULIRG MKN231 \cite{braito2004}. TW Hya is meant to be representative of a young star, a source which can illuminate protoplanetary disks and protostellar envelopes, although the spectra of such stars can vary. The spectrum of MKN231 does not correspond to a spectrum illuminating any specific molecular cloud, therefore it should only be taken as an illustration of a harder and flatter X-ray spectrum than that of a young star. We then applied to each of these spectra an attenuation corresponding to various column densities of a gas and dust mix using the attenuation cross-section of the standard model of ref \cite{bethell2011}. The elemental abundances and dust parameters of this reference are taken for protoplanetary disks. We make the assumption that these numbers should not be too different for other types of molecular gas and dust environments. The spectra considered are given in the Supplementary Information.

The integral of the yield spectrum Y$^{inc}$($\lambda$) multiplied by the normalized flux spectrum $\phi(\lambda)$ gives the average photodesorption yield \cite{dupuy2017a}:
\begin{linenomath} $$ Y^{avg} = \frac{\int Y^{inc}(\lambda)\phi(\lambda)d\lambda}{\int\phi(\lambda)d\lambda} $$ \end{linenomath}
The results for different sources and attenuations are given in table 2. These numbers can be used in astrochemical models to treat X-ray photodesorption similarly to the way UV photodesorption is treated, using the rate:
\begin{linenomath} $$ k^{PD}_i = \theta_i\sigma_{gr}n_{gr}Y^{avg}_iF_{X} \; \text{(cm$^{-3}$.s$^{-1}$)}$$ \end{linenomath}
Where $\theta_i$ is the surface coverage of species i, $\sigma_{gr}n_{gr}$ the effective grain surface and F$_X$ the integrated X-ray photon flux. 

\subsection*{Acknowledgements}

\textit{We would like to thank C. Walsh for her helpful insights on X-rays in protoplanetary disks, D. Lis for his comments on the paper and P. Marie-Jeanne for technical support. We acknowledge SOLEIL for provision of synchrotron radiation facilities under the project 20161406 and we thank N. Jaouen and the SEXTANTS team for their help on the beamline. This work was supported by the Programme National “Physique et Chimie du Milieu Interstellaire” (PCMI) of CNRS/INSU with INC/INP co-funded by CEA and CNES. Financial support from the LabEx MiChem, part of the French state funds managed by the ANR within the investissements d'avenir program under reference ANR-11-10EX-0004-02, and by the Ile-de-France region DIM ACAV program, is gratefully acknowledged. This work was done in collaboration and with financial support by the European Organization for Nuclear Research (CERN) under the collaboration agreement KE3324/TE.}

\subsection*{Author contributions}

\textit{RD treated and analysed the data, and wrote the manuscript. MB, GF, MH and JHF provided extensive input on the data analysis and the manuscript. JHF, MB, GF and RD initiated and supervised the project. JHF, MB and PJ designed the experimental set-up. GF contributed to the bibliographic work. All authors participated to the experimental runs at the SOLEIL synchrotron where the data were acquired.}

\bibliographystyle{naturemag}
\bibliography{H2O_X_astro}

\newpage
\renewcommand{\figurename}{Supplementary figure}

\section*{Supplementary information}

\subsection*{Ion desorption}
 
Ions were detected using a quadrupole mass-spectrometer. A mass spectrum of the positive ions is presented in supplementary fig. \ref{mspec}. The desorption yield of ions was calculated by first estimating the detection efficiency of the QMS in the setting used here. For this we compared the signal we obtained for C$^+$ desorption from a CO ice at 550 eV with the quantified data given in ref \cite{rosenberg1985}. From this we obtain a detection efficiency of approximately 2\% for C$^+$. The detection efficiency for other ions then depends on the mass apparatus function of the QMS, which we calibrated. Because our calibration for ions relies on data from the literature which does not have an uncertainty associated to it, the numbers should be taken with caution.

The desorption spectrum of H$^+$ shown in supplementary fig. \ref{H+} does not follow exactly the total electron yield, i.e. the ice bulk absorption. This had already been noted and investigated in detail in previous studies \cite{coulman1990, mase2000} where interpretations can be found.

\begin{figure}[b]
    \includegraphics[trim={1cm 0.5cm 1cm 1cm},clip,width=\linewidth]{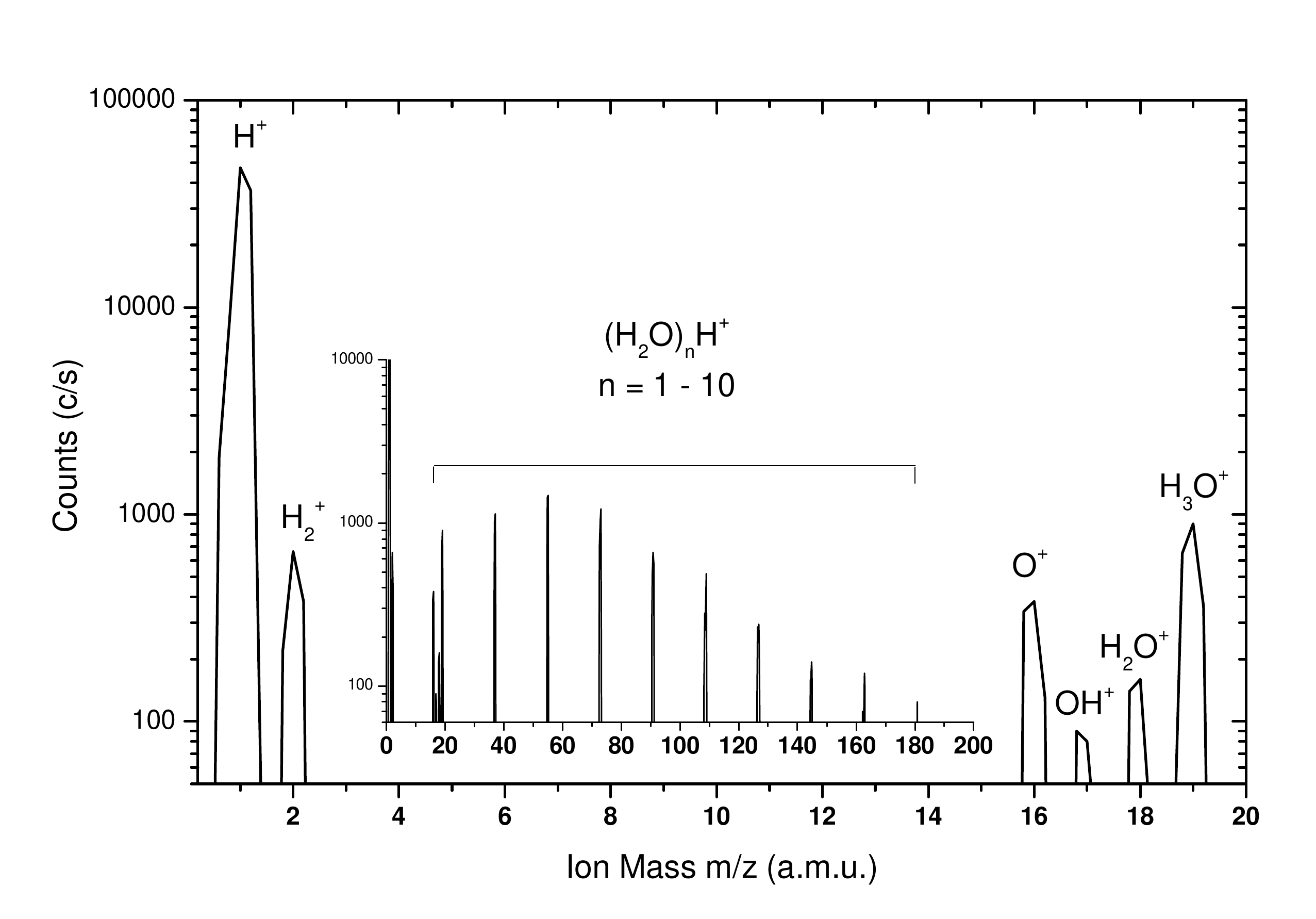}
    \caption{Mass spectrum of the desorption of positive ions by 550 eV photons. The data is for a 100 ML c-ASW ice at 90 K.}
    \label{mspec}
\end{figure}

\begin{figure}[b]
    \includegraphics[trim={0.5cm 1cm 2cm 1.5cm},clip,width=\linewidth]{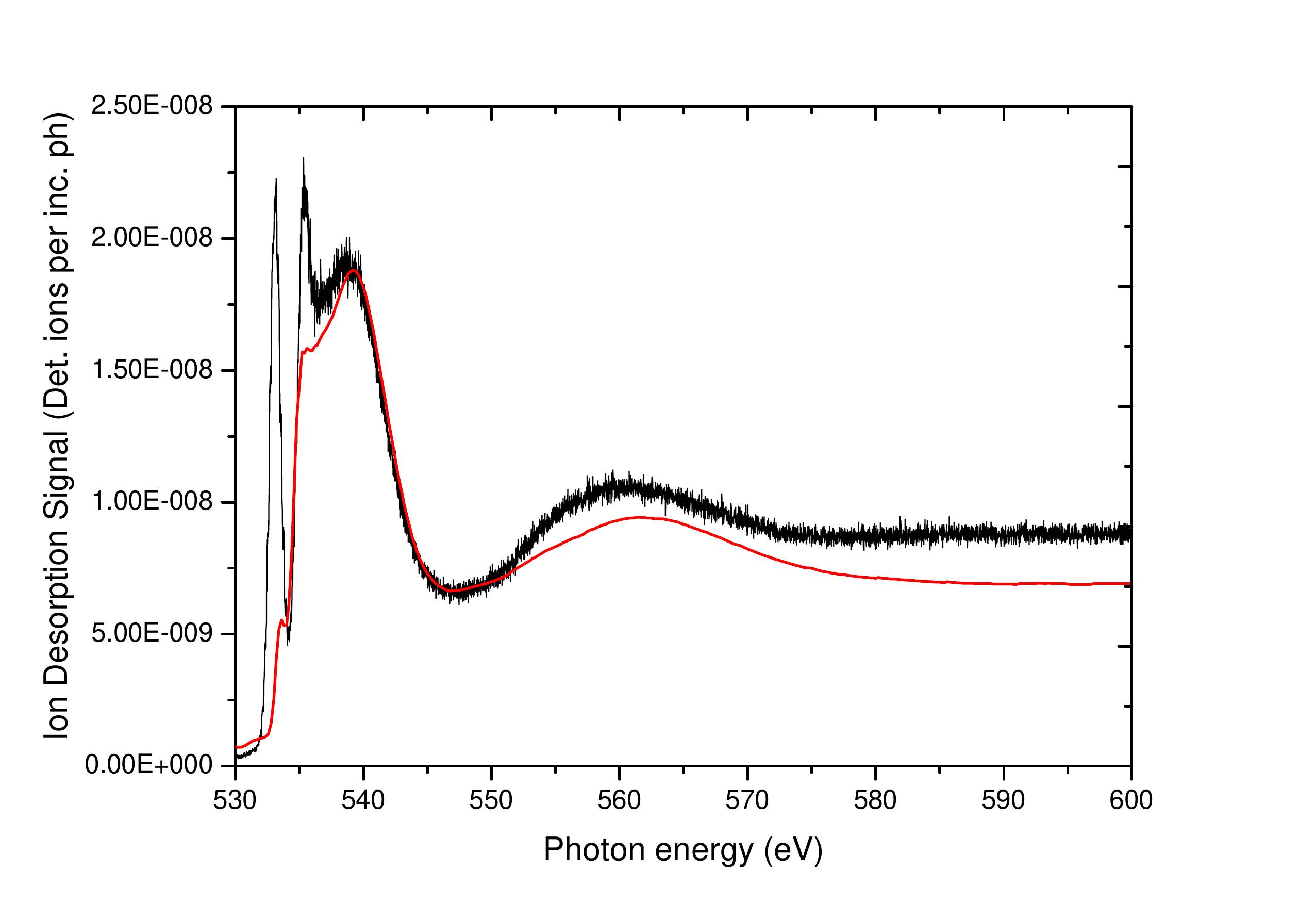}
    \caption{Photodesorption spectrum of H$^+$ (black trace). The total electron yield (red trace) was included for comparison. The data is for a 100 ML c-ASW ice at 90 K.}
    \label{H+}
\end{figure}

\subsection*{Extrapolated photodesorption yields}

\begin{figure}
    \includegraphics[trim={0.5cm 1cm 1cm 1cm},clip,width=\linewidth]{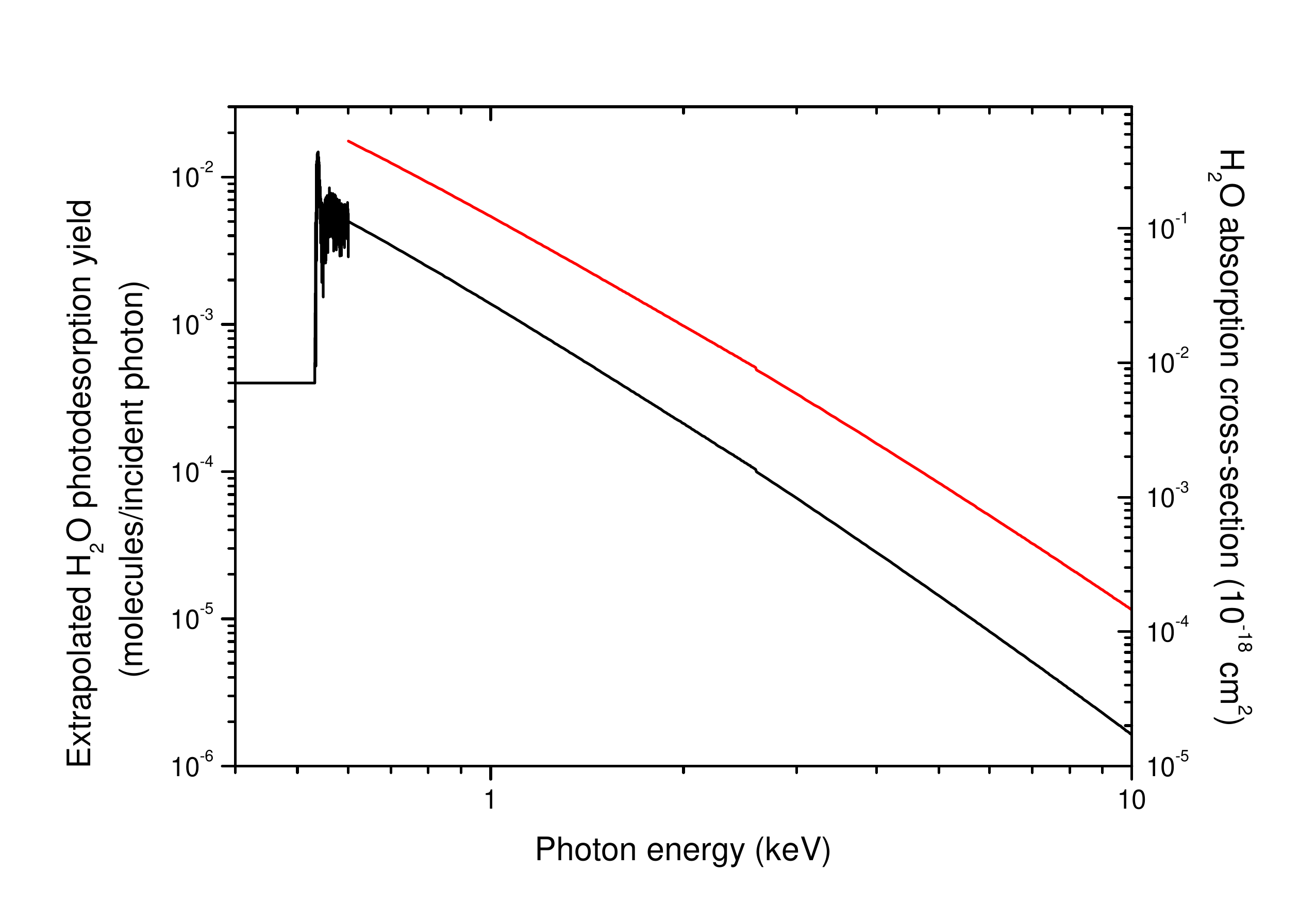}
    \caption{Photodesorption yield of neutral H$_2$O as a function of photon energy, extrapolated from 0.15 to 10 keV (black trace). The data is for a 100 ML c-ASW ice at 15K. The extrapolation is described in the Methods section of the paper. Red trace: absorption cross-section of gas-phase water obtained from \cite{berkowitz2002} and used for the extrapolation.}
    \label{yield}
\end{figure}

Supplementary fig. \ref{yield} shows the photodesorption yield of H$_2$O extrapolated from 400 to 10 000 eV using the gas phase absorption cross-section represented as well. Details are given in the Methods section of the article.

\begin{figure}
    \includegraphics[trim={1.5cm 1cm 1.5cm 0.5cm},clip,width=\linewidth]{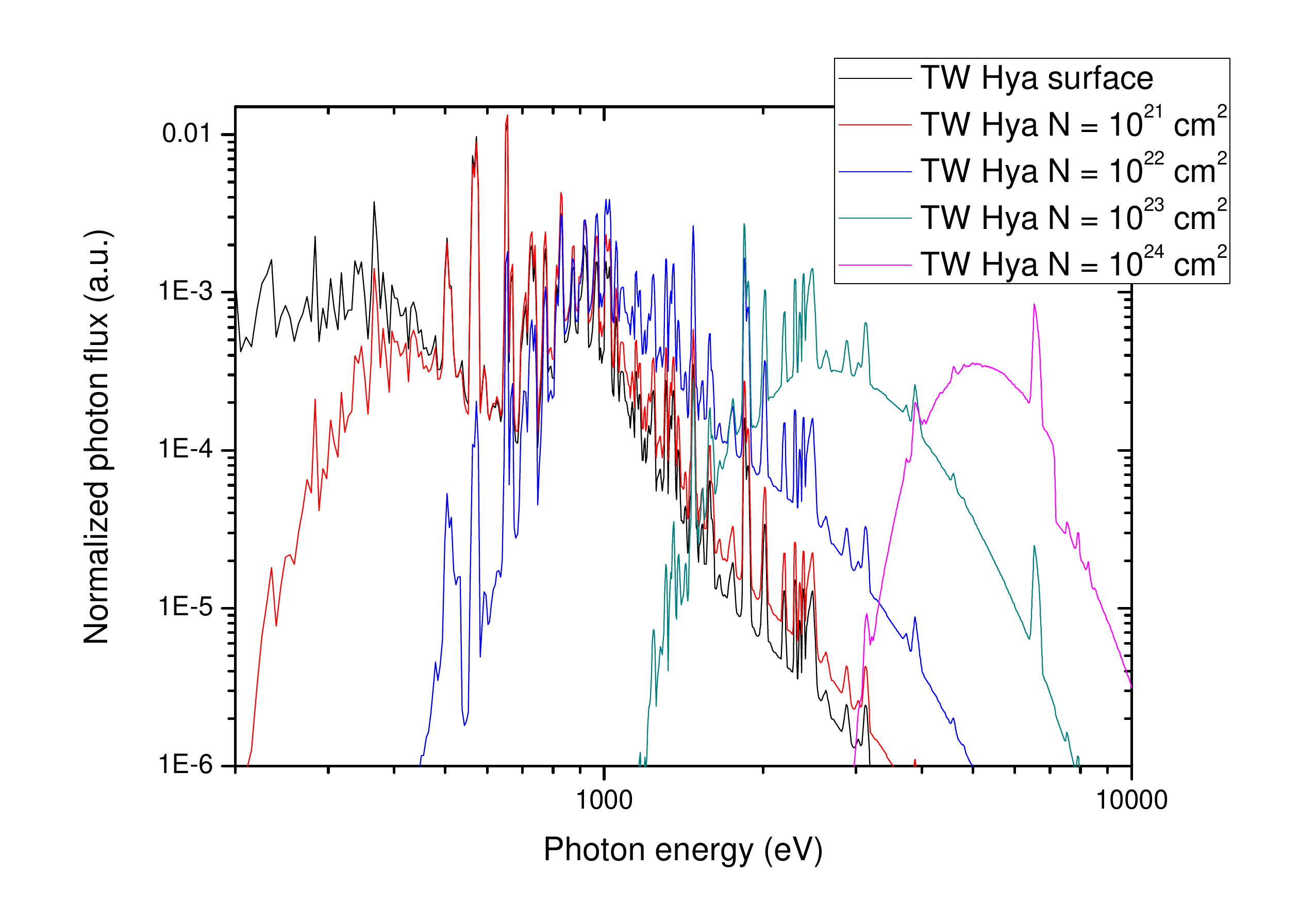}
    \caption{X-ray spectrum of TW Hya attenuated by different column densities of gas and dust. The source spectrum was retrieved from ref \cite{nomura2007} and the attenuated spectra for different column densities (N = 0 - 10$^{24}$ cm$^2$) were calculated according to the attenuation cross-sections of ref \cite{bethell2011}.}
    \label{TW}
\end{figure}

\begin{figure}
    \includegraphics[trim={1cm 1cm 1.5cm 1.5cm},clip,width=\linewidth]{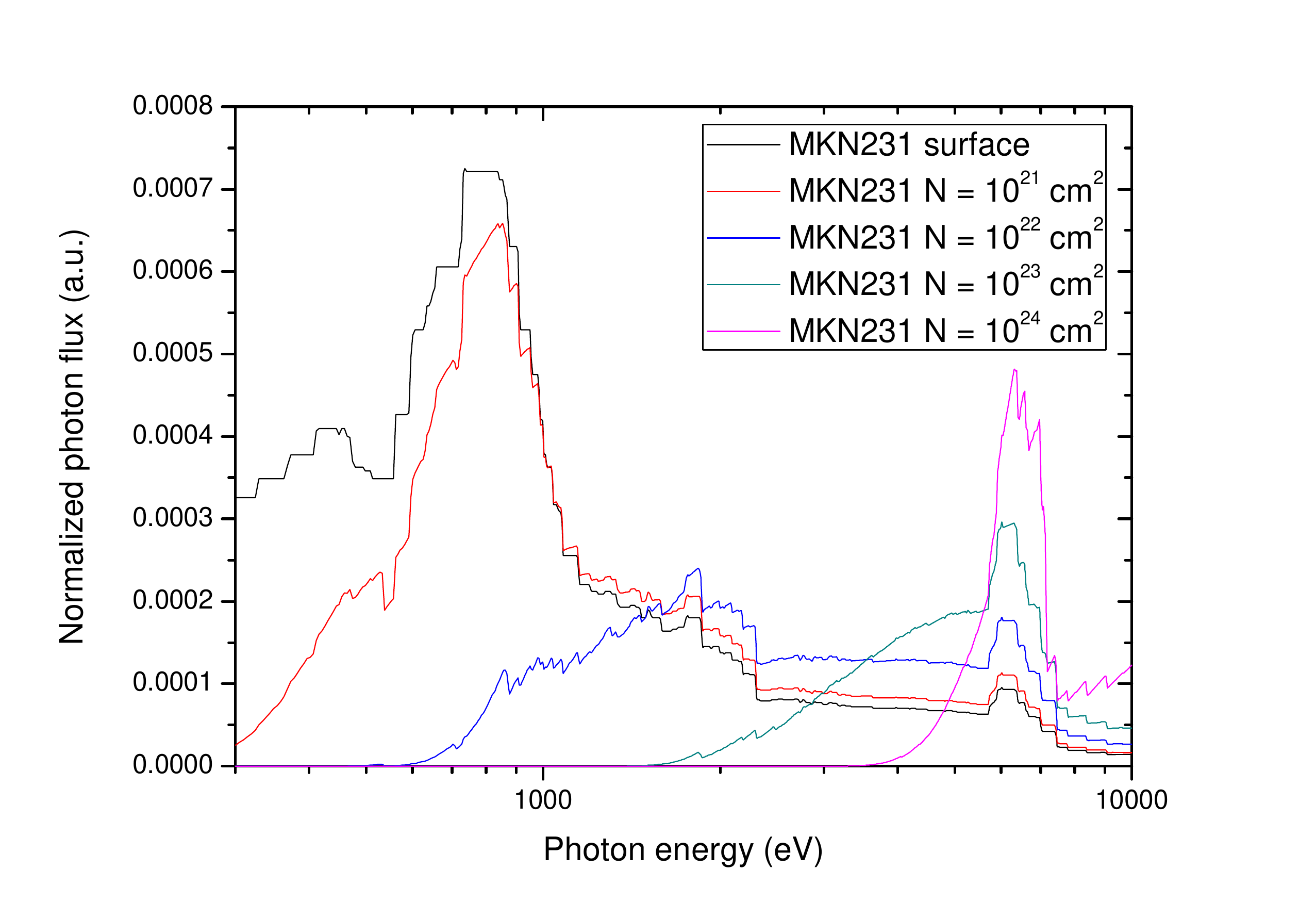}
    \caption{X-ray spectrum of the ULIRG MKN 231 attenuated by different column densities of gas and dust. The source spectrum was retrieved from ref \cite{braito2004} and the attenuated spectra for different column densities (N = 0 - 10$^{24}$ cm$^2$) were calculated according to the attenuation cross-sections of ref \cite{bethell2011}.}
    \label{MKN}
\end{figure}

Supplementary fig. \ref{TW} and \ref{MKN} show the X-ray spectra used to calculate the values given in table 2 of the text. They correspond to the X-ray spectrum of two different sources, the star TW Hya \cite{nomura2007} and the emission from the ULIRG MKN231 \cite{braito2004}, attenuated by different column densities of gas (using the cross-sections of ref \cite{bethell2011}), which displaces the peak of the spectrum towards higher and higher energies. The spectra are all normalized to their respective area for readability.

\end{document}